# Mie scattering and optical forces from evanescent fields: A complex-angle approach

**Aleksandr Y. Bekshaev,[1*] Konstantin Y. Bliokh,[2,3] and Franco Nori[2,4]**

[1]*I. I. Mechnikov National University, Dvorianska 2, Odessa, 65082, Ukraine*
[2]*Advanced Science Institute, RIKEN, Wako-shi, Saitama 351-0198, Japan*
[3]*A. Usikov Institute of Radiophysics and Electronics, 12 Ak. Proskury St., Kharkov 61085, Ukraine*
[4]*Physics Department, University of Michigan, Ann Arbor, Michigan 48109-1040, USA*
*[*]bekshaev@onu.edu.ua*

**Abstract:** Mie theory is one of the main tools describing scattering of propagating electromagnetic waves by spherical particles. Evanescent optical fields are also scattered by particles and exert radiation forces which can be used for optical near-field manipulations. We show that the Mie theory can be naturally adopted for the scattering of evanescent waves via rotation of its standard solutions by a *complex angle*. This offers a simple and powerful tool for calculations of the scattered fields and radiation forces. Comparison with other, more cumbersome, approaches shows perfect agreement, thereby validating our theory. As examples of its application, we calculate angular distributions of the scattered far-field irradiance and radiation forces acting on dielectric and conducting particles immersed in an evanescent field.

**References and links**

1. G. Mie, "Beiträge zur Optik trüber Medien, speziell kolloidaler Metallösungen," Ann. Phys. (Leipzig) **25**, 377–445 (1908).
2. M. Born and E. Wolf, *Principles of Optics*, 7th edn. (London: Pergamon, 2005).
3. H. C. Van de Hulst, *Light scattering by small particles* (New York: Chapman & Hall, 1957).
4. C. F. Bohren and D. R. Huffman, *Absorption and scattering of light by small particles* (New York: Wiley, 1983).
5. M. Dienerowitz, M. Mazilu, and K. Dholakia, "Optical manipulation of nanoparticles: a review," J. Nanophotonics **2**, 021875 (2008).
6. I. Brevik, "Experiments in phenomenological electrodynamics and the electromagnetic energy-momentum tensor," Phys. Rep. **52**, 133–201 (1979).
7. M. A. Paesler and P. J. Moyer, *Near–Field Optics* (New York: John Wiley & Sons, 1996).
8. M. Nieto-Vesperinas and N. Garcia, eds., *Optics at the nanometer scale*, NATO ASI Series (Dordrecht: Kluwer Academic Publishing, 1996).
9. S. A. Maier, *Plasmonics: Fundamentals and Applications* (New York: Springer, 2007).
10. H. Chew, D.-S. Wang, and M. Kerker, "Elastic scattering of evanescent electromagnetic waves," Appl. Opt. **18**, 2679–2687 (1979).
11. C. Liu, T. Kaiser, S. Lange, and G. Schweiger, "Structural resonances in a dielectric sphere illuminated by an evanescent wave," Opt. Commun. **117**, 521–531 (1995).
12. M. Quinten, A. Pack, and R. Wannemacher, "Scattering and extinction of evanescent waves by small particles," Appl. Phys. B **68**, 87–92 (1999).
13. E. Almaas and I. Brevik, "Radiation forces on a micrometer-sized sphere in an evanescent field," J. Opt. Soc. Am. B **12**, 2429–2438 (1995).
14. I. Brevik, T. A. Sivertsen, and E. Almaas, "Radiation forces on an absorbing micrometer-sized sphere in an evanescent field," J. Opt. Soc. Am. B **20**, 1739–1749 (2003).
15. H. Y. Jaising and O. G. Hellesø, "Radiation forces on a Mie particle in the evanescent field of an optical waveguide," Opt. Commun. **246**, 373–383 (2005).
16. S. Chang, J. T. Kim, J. H. Jo, and S. S. Lee, "Optical force on a sphere caused by the evanescent field of a Gaussian beam; effects of multiple scattering," Opt. Commun. **139**, 252–261 (1997).

17. Y. G. Song, B. M. Han, and S. Chang, "Force of surface plasmon-coupled evanescent fields on Mie particles," Opt. Commun. **198**, 7–19 (2001).
18. J. Y. Walz, "Ray optics calculation of the radiation forces exerted on a dielectric sphere in an evanescent field," Appl. Opt. **38**, 5319–5330 (1999).
19. P.C. Chaumet and M. Nieto-Vesperinas, "Electromagnetic force on a metallic particle in the presence of a dielectric surface," Phys. Rev. B **62**, 11185–11191 (2000).
20. J. R. Arias-González and M. Nieto-Vesperinas, "Optical forces on small particles: attractive and repulsive nature and plasmon-resonance conditions," J. Opt. Soc. Am. A **20**, 1201–1209 (2003).
21. M. Nieto-Vesperinas, P. C. Chaumet, and A. Rahmani, "Near-field photonic forces," Phil. Trans. R. Soc. Lond. A **362**, 719–737 (2004).
22. M. Nieto-Vesperinas and J. J. Saenz, "Optical forces from an evanescent wave on a magnetodielectric small particle," Opt. Lett. **35**, 4078–4080 (2010).
23. M. Nieto-Vesperinas and J. R. Arias-González, "Theory of forces induced by evanescent fields," *arXiv*: 1102.1613 (2011).
24. S. Kawata and T. Sugiura, "Movement of micrometer–sized particles in the evanescent field of a laser beam," Opt. Lett. **17**, 772–774 (1992).
25. S. Kawata and T. Tani, "Optically driven Mie particles in an evanescent field along a channeled waveguide," Opt. Lett. **21**, 1768–1770 (1996).
26. M. Vilfan, I. Muševič, and M. Čopič, "AFM observation of force on a dielectric sphere in the evanescent field of totally reflected light," Europhys. Lett. **43**, 41–46 (1998).
27. K. Sasaki, J.-I. Hotta, K.-I. Wada, and H. Masuhara, "Analysis of radiation pressure exerted on a metallic particle within an evanescent field," Opt. Lett. **25**, 1385–1387 (2000).
28. L. N. Ng, M. N. Zervas, J. S. Wilkinson, and B. J. Luff, "Manipulation of colloidal gold nanoparticles in the evanescent field of a channel waveguide," Appl. Phys. Lett. **76**, 1993–1995 (2000).
29. G. Volpe, R. Quidant, G. Badenes, and D. Petrov, "Surface plasmon radiation forces," Phys. Rev. Lett. **96**, 238101 (2006).
30. M. Šiler, T. Čižmár, M. Šerý, and P. Zemánek, "Optical forces generated by evanescent standing waves and their usage for sub-micron particle delivery," Appl. Phys. B **84**, 157–165 (2006).
31. S. Gaugiran, S. Getin, J. M. Fedeli, and J. Derouard, "Polarization and particle size dependence of radiative forces on small metallic particles in evanescent optical fields. Evidences for either repulsive or attractive gradient forces," Opt. Express **15**, 8146–8156 (2007).
32. D. C. Prieve and J. Y. Walz, "Scattering of an evanescent surface wave by a microscopic dielectric particle," Appl. Opt. **32**, 1629–1641 (1993).
33. A. Bekshaev, K. Bliokh, and M. Soskin, "Internal flows and energy circulation in light beams," J. Opt. **13**, 053001 (2011).
34. K. Y. Bliokh and F. Nori, "Transverse spin of a surface polariton," Phys. Rev. A **85**, 061801(R) (2012).
35. F. I. Fedorov, "To the theory of total reflection," Dokl. Akad. Nauk SSSR **105**, 465–468 (1955) [reprinted in J. Opt. **15**, 014002 (2013)].
36. C. Imbert, "Calculation and experimental proof of the transverse shift induced by total internal reflection of a circularly polarized light beam," Phys. Rev. D **5**, 787–796 (1972).
37. S. H. Simpson and S. Hanna, "Orbital motion of optically trapped particles in Laguerre–Gaussian beams," J. Opt. Soc. Am. A **27**, 2061–2071 (2010).
38. A. Y. Bekshaev, O. V. Angelsky, S. G. Hanson, and C. Y. Zenkova, "Scattering of inhomogeneous circularly polarized optical field and mechanical manifestation of the internal energy flows," Phys. Rev. A **86**, 023847 (2012).

## 1. Introduction

The scattering of light by various particles appears in a variety of optical processes, with applications ranging from microscopy to astrophysics. A fundamental solution describing electromagnetic wave scattering by a spherical particle was found in 1908 by Gustav Mie [1]. Since then, the *Mie theory* has become the main tool for characterization of particle-induced light scattering [2–4]. In addition to the properties of scattered light, this theory allows calculation of *radiation forces* exerted on particles. Such forces are of great importance for optical manipulations [5] and for investigations of the fundamental physical properties of electromagnetic fields [6].

With the development of near-field optics and plasmonics [7–9], *evanescent* electromagnetic waves have attracted enormous interest, both for theory and applications. In particular, the evanescent-wave scattering by small particles and accompanying radiation forces are important in modern optics. Analogues of the Mie theory for evanescent waves were elaborated [10–13], while the radiation forces from evanescent fields were extensively examined theoretically [13–23] and experimentally [24–31]. Most of these Mie-type approaches are based on a straightforward expansion of the incident evanescent wave in a series of vector spherical harmonics and the subsequent reproduction of the Mie procedure, which results in rather cumbersome calculations [10–17]. Alternatively, one can treat analytically the simplest approximation of the dipole Rayleigh scattering of evanescent waves by small (much less than the wavelength) particles [19–23].

In this paper we put forward a much simpler method for the calculation of the Mie scattering and optical forces with an evanescent incident field. Our approach uses the fact that an evanescent wave can be represented as a regular $z$-propagating plane wave which is geometrically rotated by a *complex angle*. Therefore, using the rotational symmetry of the Mie problem, we find that the scattered field can also be obtained by applying the same complex-angle rotation to the *well-known* Mie-theory solutions. In other words, one does not need to solve a new problem but only apply a simple geometric transformation to the known solutions. Naturally, our results coincide with those obtained by previous exact methods [10–15] or in the dipole-scattering approximation for small particles [19–23]. However, our approach offers significant advantages, including a much more transparent and time-saving procedure, use of the well-elaborated Mie-theory calculation schemes, existing software codes, etc.

To demonstrate several applications of our method, we first calculate the far-field scattering diagrams for different evanescent-wave polarizations and particle sizes, and then compare them with the usual Mie-theory diagrams for the incident plane wave case. Second, we compute optical forces exerted on dielectric particles in an evanescent field from a totally-reflecting interface, and show that our results coincide with those previously reported in [13,22,23]. Finally, we calculate the optical force on metallic particles and address the problem of the positive vertical force repelling the particle from the surface [14,15,19,20,23,24,31]. Note that, as in most other works, we neglect multiple reflections from the surface limiting the evanescent field. More accurate treatments [16,17,19–21,23,32] show that the influence of these reflections can be neglected in a wide range of parameters: e.g., in calculations of the parallel components of the force, and for particle sizes of the order of the wavelength and not exhibiting resonances. Even in cases where multiple-scattering effects must be taken into account, the single scattering of a pure evanescent wave is the first step, and it can be facilitated by the approach proposed here.

## 2. Incident field configuration

The standard formulation of the Mie scattering theory starts with an incident monochromatic plane wave propagating along the $z$ axis, with wavevector $\mathbf{k} = (0,0,k)$, whereas the center of the particle is located at the origin $(x,y,z) = \mathbf{0}$. Complex electric and magnetic field amplitudes of the incident wave are written as[1]

$$\mathbf{E} = \begin{pmatrix} E_\parallel \\ E_\perp \\ 0 \end{pmatrix} \exp(ikz), \quad \mathbf{H} = \begin{pmatrix} H_\parallel \\ H_\perp \\ 0 \end{pmatrix} \exp(ikz) = \sqrt{\frac{\varepsilon}{\mu}} \begin{pmatrix} -E_\perp \\ E_\parallel \\ 0 \end{pmatrix} \exp(ikz). \qquad (1)$$

Here the subscripts $\parallel$ and $\perp$ denote the $p$- and $s$-polarizations with respect to the $(x,z)$ plane, $k = \omega n/c \equiv n k_0$ is the wave number [$\omega$ is the frequency, $c$ is the velocity of light in vacuum, and throughout the paper we omit the common factor $\exp(-i\omega t)$], and we assume a lossless medium characterized by permittivity $\varepsilon$, permeability $\mu$, and refractive index $n = \sqrt{\varepsilon\mu}$.

---

[1] In this paper we use the Gaussian system of units. For transition to the SI units, one should modify the field amplitudes as $\mathbf{E} \to \sqrt{\varepsilon_0}\mathbf{E}$, $\mathbf{H} \to \sqrt{\mu_0}\mathbf{H}$ ($\varepsilon_0$ and $\mu_0$ are the vacuum permittivity and permeability, $\varepsilon_0 \mu_0 = c^{-2}$), and use the corresponding constant $g = 1/2$ [see Eq. (8) below].

Let us consider a more generic situation, where an incident wave propagates at some angle $\gamma$ with respect to the $z$ axis in the $(x,z)$ plane. The wave field is obtained using the corresponding rotation operator $\hat{\mathcal{R}}_y(-\gamma) = \exp(-i\gamma \hat{J}_y / \hbar)$, where $\hat{\mathbf{J}}$ is the total (spin plus orbital) angular momentum operator. Application of the operator $\hat{\mathcal{R}}_y(-\gamma)$ (which rotates both vector directions and function distributions) to the wave fields $\mathbf{E}(\mathbf{r})$ and $\mathbf{H}(\mathbf{r})$ results in the transformation:

$$\mathbf{E}(\mathbf{r}) \to \hat{R}_y(-\gamma) \mathbf{E}\left[\hat{R}_y(\gamma)\mathbf{r}\right], \quad \mathbf{H}(\mathbf{r}) \to \hat{R}_y(-\gamma) \mathbf{H}\left[\hat{R}_y(\gamma)\mathbf{r}\right], \quad (2)$$

where

$$\hat{R}_y(-\gamma) = \begin{pmatrix} \cos\gamma & 0 & \sin\gamma \\ 0 & 1 & 0 \\ -\sin\gamma & 0 & \cos\gamma \end{pmatrix} \quad (3)$$

is the rotation matrix which acts on the Cartesian components of the vectors. Explicitly, Eqs. (2) and (3) yield

$$\mathbf{E} = \begin{pmatrix} E_\parallel \cos\gamma \\ E_\perp \\ -E_\parallel \sin\gamma \end{pmatrix} \exp\left[ik(z\cos\gamma + x\sin\gamma)\right],$$

$$\mathbf{H} = \sqrt{\frac{\varepsilon}{\mu}} \begin{pmatrix} -E_\perp \cos\gamma \\ E_\parallel \\ E_\perp \sin\gamma \end{pmatrix} \exp\left[ik(z\cos\gamma + x\sin\gamma)\right]. \quad (4)$$

Here the transformation of coordinates $\mathbf{r} \to \hat{R}_y(\gamma)\mathbf{r}$ is equivalent to the wavevector rotation $\mathbf{k} \to \hat{R}_y(-\gamma)\mathbf{k}$.

Importantly, the simple expressions (2)–(4), which describe an obliquely-propagating plane wave, can also describe *evanescent* plane waves decaying away from the $z = 0$ plane (Fig. 1). Indeed, consider now the *complex* propagation angle $\gamma$ given by:

$$\gamma = \frac{\pi}{2} - i\alpha, \quad \alpha > 0. \quad (5)$$

In this case, the rotation matrix (3) $\hat{R}_y(-\gamma)$ takes the form

$$\hat{R}_y\left(-\frac{\pi}{2} + i\alpha\right) = \begin{pmatrix} i\sinh\alpha & 0 & \cosh\alpha \\ 0 & 1 & 0 \\ -\cosh\alpha & 0 & i\sinh\alpha \end{pmatrix}, \quad (6)$$

and the incident field is obtained by the corresponding modifications of Eqs. (4):

$$\mathbf{E} = \begin{pmatrix} iE_\parallel \sinh\alpha \\ E_\perp \\ -E_\parallel \cosh\alpha \end{pmatrix} \exp(ikx\cosh\alpha - kz\sinh\alpha),$$

$$\mathbf{H} = \sqrt{\frac{\varepsilon}{\mu}} \begin{pmatrix} -iE_\perp \sinh\alpha \\ E_\parallel \\ E_\perp \cosh\alpha \end{pmatrix} \exp(ikx\cosh\alpha - kz\sinh\alpha), \tag{7}$$

with wavevector $\mathbf{k} = (k\cosh\alpha, 0, ik\sinh\alpha)$. Equations (7) describe an evanescent plane wave propagating in the $x$-direction and decaying in the positive $z$-direction, see Fig. 1.

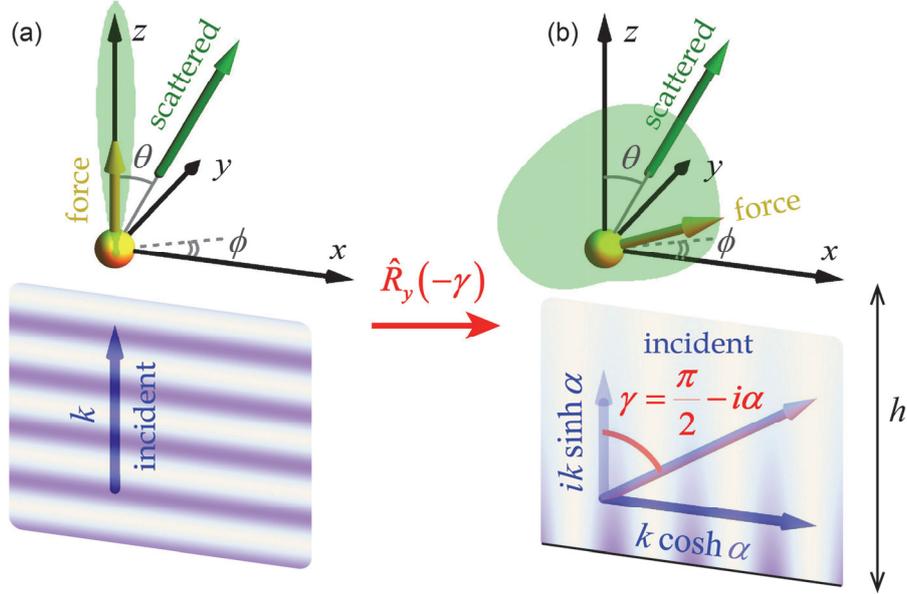

**Fig. 1.** Schematic of the Mie scattering problem. Incident wave (blue), scattered field (green), and radiation force exerted on the particle (yellow) are shown. (a) standard Mie theory with the incident plane wave propagating along the $z$-axis. (b) Rotation of the field, Eqs. (2) and (5), by the complex angle $\gamma = \frac{\pi}{2} - i\alpha$ results in the modified Mie problem with evanescent incident wave (6). The parameter $h$ indicates the distance to the surface where the evanescent wave is generated.

It is useful to consider the time-averaged densities of the electromagnetic energy and momentum in the evanescent field (7). They are determined by the well-known relations [2]

$$w = \frac{g}{2}\left(\varepsilon|\mathbf{E}|^2 + \mu|\mathbf{H}|^2\right), \quad \mathbf{p} = \frac{g}{c}\operatorname{Re}(\mathbf{E}^* \times \mathbf{H}), \tag{8}$$

where $g = (8\pi)^{-1}$ in the Gaussian system of units ($g = 1/2$ in SI units). Substituting the evanescent fields (7) into Eqs. (8), we obtain

$$w = \varepsilon g \cosh^2\alpha \left(|E_\parallel|^2 + |E_\perp|^2\right)\exp(-2kz\sinh\alpha),$$

$$p_x = \frac{gn}{c\mu}\cosh\alpha\left(|E_\parallel|^2 + |E_\perp|^2\right)\exp(-2kz\sinh\alpha), \quad p_z = 0,$$

$$p_y = -2\frac{gn}{c\mu}\sinh\alpha\cosh\alpha\,\operatorname{Im}(E_\parallel^* E_\perp)\exp(-2kz\sinh\alpha). \tag{9}$$

As expected for an evanescent field, the $z$-directed momentum component vanishes and the energy flows parallel to the $z = 0$ plane. It is worth noticing that the $x$ component of the momentum transports the energy and can be written as $p_x = v_g w/c^2$. Here $v_g = ck_0/k_x$ ($k_x = k \cosh \alpha > k$) is the wave group velocity, and $p_x$ is essentially combined from *orbital* and *spin* contributions as described in [33,34]. At the same time, the transverse momentum $p_y$ was described by Fedorov and Imbert in the total internal reflection problem [35,36]. This transverse momentum $p_y$ is proportional to the ellipticity of the wave polarization $2\,\mathrm{Im}\left(E_\parallel^* E_\perp\right)$, it is of purely *spin* nature, and therefore does not transport energy [33,34].

In practice, one of the standard ways to generate the incident evanescent wave (7) is to use the total internal reflection. For instance, let the plane $z = -h$ be the dielectric interface separating two media, with parameters $\varepsilon_1$, $\mu_1$, $n_1 = \sqrt{\varepsilon_1 \mu_1}$ ($z < -h$) and $\varepsilon$, $\mu$, $n = \sqrt{\varepsilon \mu} < n_1$ ($z > -h$). A propagating plane wave $\mathbf{E}_0$ with the $p$- and $s$-polarized field components $E_{0\parallel}$ and $E_{0\perp}$ impinges on the interface from the $z < -h$ half-space at an angle of incidence $\theta_1$, such that the condition for total internal reflection $n_1 \sin \theta_1 > n$ is realized. Then, the transmitted field at $z > -h$ calculated via the corresponding Snell-Fresnel formulae [2] will be the evanescent wave (7) with parameters

$$\cosh \alpha = \frac{n_1}{n} \sin \theta_1, \quad \sinh \alpha = \sqrt{\left(\frac{n_1}{n}\right)^2 \sin^2 \theta_1 - 1}, \quad (10)$$

$$E_\parallel = \frac{2 \frac{n}{n_1} \cos \theta_1}{\frac{\varepsilon}{\varepsilon_1} \cos \theta_1 + i \frac{n}{n_1} \sinh \alpha} e^{-kh \sinh \alpha} E_{0\parallel}, \quad E_\perp = \frac{2 \frac{\mu}{\mu_1} \cos \theta_1}{\frac{\mu}{\mu_1} \cos \theta_1 + i \frac{n}{n_1} \sinh \alpha} e^{-kh \sinh \alpha} E_{0\perp}. \quad (11)$$

One can notice that the complex angle of rotation $\gamma$, determined by Eqs. (5) and (10), is just the complex angle of refraction which formally follows from Snell's law under the total-reflection conditions [2]. Note also that the evanescent wave generated at the distance $h$ from the $z = 0$ plane acquires the amplitude attenuation factor $\exp(-kh \sinh \alpha)$.

## 3. Complex-angle Mie theory: Scattered field and radiation forces

Considering light scattering by a spherical particle of radius $a$ with electromagnetic parameters $\varepsilon_p$, $\mu_p$, and $n_p = \sqrt{\varepsilon_p \mu_p}$, the standard Mie equations establish linear relation between the amplitudes of the incident plane wave (1), $\mathbf{E}$, $\mathbf{H}$ and scattered fields $\mathbf{E}^s$, $\mathbf{H}^s$. These known equations [4] are collected in Appendix A in a complete form, keeping the scattered near field and radial components (which are typically omitted). One can write these equations in a symbolic operator form as

$$\mathbf{E}^s(\mathbf{r}) = \hat{S}_E(\mathbf{r}) \mathbf{E}, \quad \mathbf{H}^s(\mathbf{r}) = \hat{S}_H(\mathbf{r}) \mathbf{E}, \quad (12)$$

where the Mie scattering operators $\hat{S}_{E,H}(\mathbf{r})$ are well defined in the complex domain.

We have shown that the rotation (6) by a complex angle transforms the incident $z$-propagating plane wave (1) to the evanescent wave (7). Can we then apply a similar transformation to the whole Mie scattering problem (12)? This problem is linear. The free-space Maxwell equations for the complex field amplitudes and the boundary conditions (for the case of a spherical surface centered at the origin) are both invariant with respect to rotations. Hence, *the solution of the Mie problem with the incident evanescent wave (7) can be obtained by applying the complex-angle rotation (6) to the standard Mie solution (12) with the incident plane wave:*

$$\mathbf{E}^s(\mathbf{r}) \to \hat{R}_y\left(-\frac{\pi}{2} + i\alpha\right) \mathbf{E}^s\left[\hat{R}_y\left(\frac{\pi}{2} - i\alpha\right) \mathbf{r}\right],$$

$$\mathbf{H}^s(\mathbf{r}) \to \hat{R}_y\left(-\frac{\pi}{2}+i\alpha\right)\mathbf{H}^s\left[\hat{R}_y\left(\frac{\pi}{2}-i\alpha\right)\mathbf{r}\right]. \tag{13}$$

The total field is given by the vector summation of the incident and scattered fields:

$$\mathbf{E}^{tot} = \mathbf{E}+\mathbf{E}^s, \quad \mathbf{H}^{tot} = \mathbf{H}+\mathbf{H}^s. \tag{14}$$

The fairly simple equations (6), (7), (12)–(14) (supplemented with the standard Mie formulas in Appendix A) represent the central results of this paper, namely, the *complex-angle Mie theory* for evanescent incident waves. This theory is mathematically equivalent to the previous exact methods [10–15] based on the explicit expansion of the incident evanescent field in spherical functions and the boundary-problem solution (we show this in Section 4). However, our method is free from tedious analytic transformations and the whole calculation requires only the *standard* Mie formulas derived for the *z*-directed incident plane wave. Hence, it offers considerable advantages, such as well-developed theoretical approaches, elaborated calculation schemes, and available computer software codes.

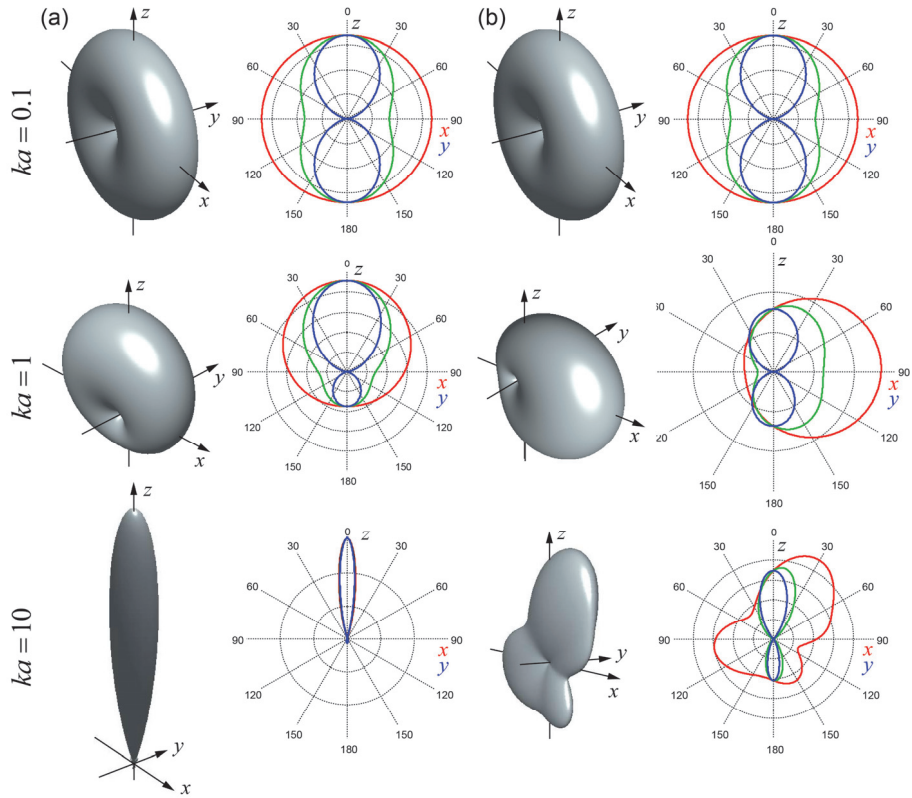

**Fig. 2.** Angular diagrams for the scattering far-field irradiance $I^s(\theta,\phi) = \text{Re}\left[\mathbf{E}^{s*}(\theta,\phi)\times\mathbf{H}^s(\theta,\phi)\right]_r$ at $r \gg \max(a,\lambda)$ for the *s*-polarized incident field, dielectric particle with $m = n_p/n = 1.75$, and for different particle sizes. (a) Standard Mie scattering with incident plane wave (1) and (b) complex-angle Mie scattering for the incident evanescent wave (7) with $\sinh\alpha = 0.92$. Red, green, and blue curves in the polar plots represent cross-sections of the 3D scattering diagrams by azimuthal planes with $\phi = 0$ $(x,z)$, $\phi = \pi/4$, and $\phi = \pi/2$ $(x,y)$, respectively. Diagrams in the (a) and (b) panels are related to each other via the complex-angle rotation (13).

To demonstrate the ability of our method to characterize the Mie scattering of evanescent waves using the same code as for the standard Mie theory, in Figures 2 and 3 we plot angular diagrams of the irradiance of the scattering far-fields (12) and (13) for different polarizations and particle sizes. One can see there that for small particles, $ka \ll 1$ (the dipole Rayleigh scattering case), the scattering indicatrix is entirely similar for the *s*-polarized propagating and evanescent incident waves (Fig. 2), whereas for larger particles, $ka \geq 1$, the evanescent-wave scattering acquires a natural $x \leftrightarrow -x$ asymmetry. For the *p*-polarized incident fields (Fig. 3), the scattering patterns for the propagating and evanescent incident waves differ from each other even for small dipole particles. This is because a *p*-polarized evanescent field inevitably possesses a non-zero *z*-component, see Eq. (7). It should be noticed that the far-field scattering diagrams have somewhat restricted meaning for the incident evanescent field because of the inevitable presence of an interface bounding the evanescent wave. Nonetheless, they can indicate characteristic features of the scattered field, particularly in the directions parallel to the interface.

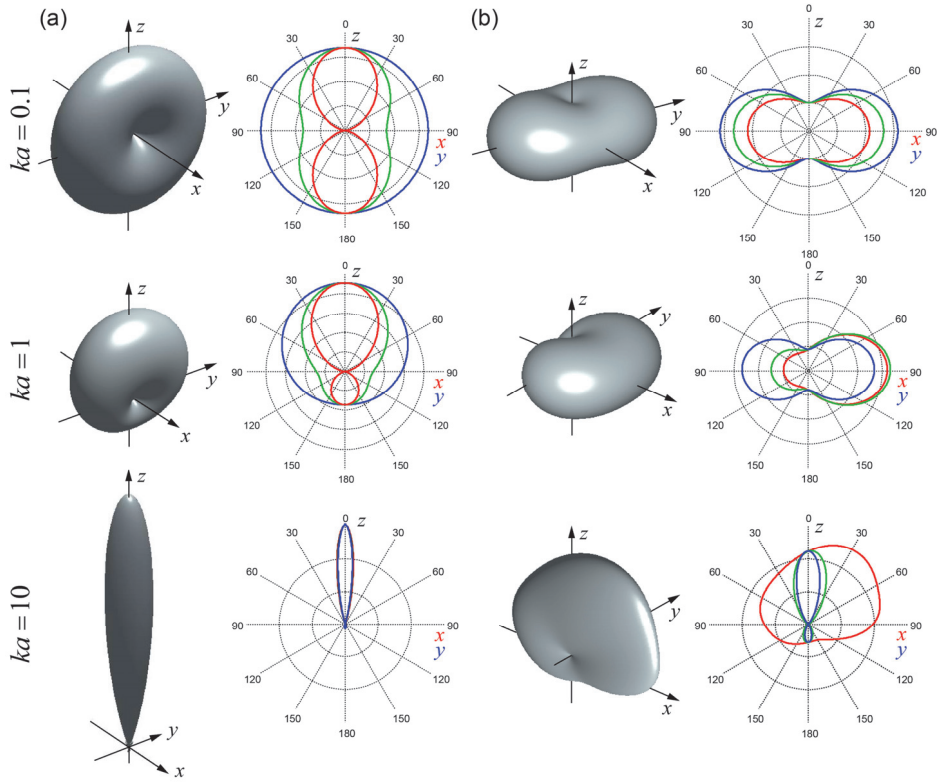

**Fig. 3.** Same as in Fig. 2, but for the *p*-polarized incident waves.

One of the important applications of the Mie theory is the calculation of the optical force exerted on a particle by the total electromagnetic field (14). This force is determined by the Maxwell stress tensor $\hat{T} = \{T_{ij}\}$, $i, j = x, y, z$:

$$T_{ij} = g \operatorname{Re}\left[ \varepsilon E_i^{tot*} E_j^{tot} + \mu H_i^{tot*} H_j^{tot} - \frac{1}{2}\delta_{ij}\left( \varepsilon \left|\mathbf{E}^{tot}\right|^2 + \mu \left|\mathbf{H}^{tot}\right|^2 \right) \right]. \tag{15}$$

Integrating the stress tensor components over any surface *A* enclosing the particle (e.g., a sphere $S = \{r = R\}$, $R > a$), we obtain the optical force:

$$\mathbf{F} = \oint_A \hat{T}\mathbf{n}\, dA = R^2 \int_S \hat{T}\mathbf{n}\, d\Omega, \tag{16}$$

where $d\Omega = \sin\theta\, d\theta\, d\phi$ is the elementary solid angle, $\mathbf{n} = (\sin\theta\cos\phi, \sin\theta\sin\phi, \cos\theta)^T$ is the unit vector of the outer normal to the sphere surface. Below we calculate the optical force from the evanescent incident field using the complex-angle Mie theory, and compare these results with the results of previous, more cumbersome, approaches.

## 4. Radiation forces: Comparison with other approaches and new applications

To verify the validity of our method, we apply the complex-angle Mie theory to problems involving optical forces from evanescent fields. We assume that a spherical particle of radius $a$ lies on the totally-reflecting surface $z = -a$ [13,18], so that its center is positioned at $z = 0$, and the incident evanescent wave is described by Eqs. (7), (10), and (11) with $h = a$.

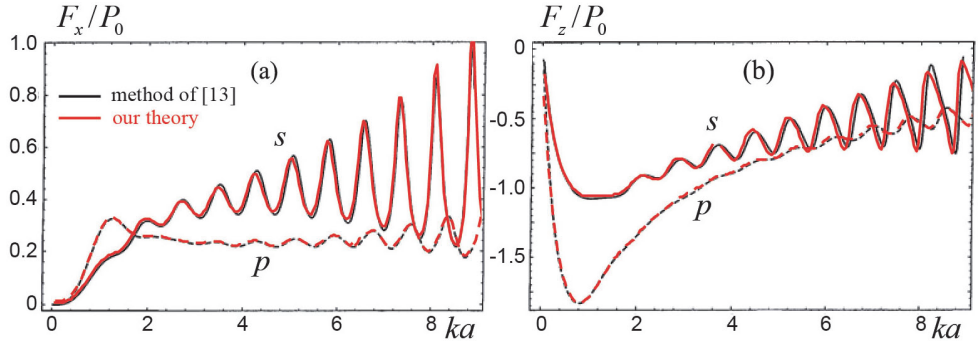

**Fig. 4.** Dimensionless radiation force components $F_{x,z}/P_0$ versus the particle-size parameter $ka$ for a dielectric particle lying on a total-reflecting surface. The parameters of the particle and incident field are given by Eqs. (7), (10), (11), and (17). The cases of the *s*-polarization (solid lines) and *p*-polarization (dashed lines) are shown. The results of our calculations based on the complex-angle Mie theory (12)–(16) (red curves) are superimposed over the data taken from Figs. 8 and 9 of Ref. [13] (black curves).

First, following the well-established approach of Refs. [13,14], we consider a dielectric particle and either *p*-polarized ($E_{0\perp} = 0$) or *s*-polarized ($E_{0\|} = 0$) incident wave. The input parameters possess the following numerical values [13]:

$$\mu_1 = \mu = \mu_p = 1, \quad n_1 = 1.75, \quad n = 1, \quad n_p = 1.5, \quad \theta_1 = 51°. \tag{17}$$

(Recall that parameters with the subscript "1", without subscript, and with subscript "*p*" correspond to the high-index medium, low-index medium, and particle, respectively.). Using Eqs. (10), this yields $\sinh\alpha \simeq 0.92$. The calculated force (16) will be normalized by

$$P_0 = \frac{a^2}{4\pi}\left(|E_{0\perp}|^2 + |E_{0\|}|^2\right). \tag{18}$$

This quantity is proportional to the time-average momentum flux of the incident plane wave $\mathbf{E}_0$ through the area $a^2$ and represents the Gaussian-unit counterpart to the SI-unit normalization divider $\varepsilon_0 a^2 E_0^2$ used in [13,14,18]. Figure 4 shows the optical force components $F_x$ and $F_z$ versus the particle size parameter $ka$, calculated using the complex-angle Mie theory (6), (7), (10)–(16), i.e., via numerical evaluation of the scattered field (12), (13) and the force integral (16). These data are superimposed over the data obtained in [13] using a considerably more complex theory. Evidently, there is an excellent agreement between the two approaches. (Small deviations can be attributed to the accuracy of numerical calculations and to graphic distortions in the printed copy of [13].)

Second, we compare the results of our calculations of radiation forces for small particles ($ka \lesssim 0.1$) with what follows from the known Rayleigh-scattering formulae for evanescent fields. From the equations derived in [21–23], it follows that for non-magnetic media and particles ($\mu_1 = \mu = \mu_p = 1$) the optical force components are given by

$$F_x = \frac{1}{2}\left[|E_\perp|^2 + |E_\parallel|^2\left(2\cosh^2\alpha - 1\right)\right] ka^3 \cosh\alpha$$
$$\times \left[\mathrm{Im}\left(\frac{\varepsilon_p - \varepsilon}{\varepsilon_p + 2\varepsilon}\right) + \frac{2}{3}(ka)^3\left|\frac{\varepsilon_p - \varepsilon}{\varepsilon_p + 2\varepsilon}\right|^2\right] e^{-2kh\sinh\alpha}, \quad (19)$$

$$F_z = -\frac{1}{2}\left[|E_\perp|^2 + |E_\parallel|^2\left(2\cosh^2\alpha - 1\right)\right] ka^3 \sinh\alpha \,\mathrm{Re}\left(\frac{\varepsilon_p - \varepsilon}{\varepsilon_p + 2\varepsilon}\right) e^{-2kh\sinh\alpha}. \quad (20)$$

The same expressions can be derived in the first electric-dipole-scattering approximation [i.e., keeping only the terms with the coefficient $a_1$ of Eq. (A4)] of our complex-angle Mie theory (12)–(16). Taking the numerical values of the parameters from [23]:

$$n_1 = 1.5, \quad n = 1, \quad \theta_1 = 42°, \quad \lambda = 632.8 \text{ nm}, \quad a = 10 \text{ nm} \ (ka \simeq 0.1), \quad (21)$$

(which yields $\sinh\alpha \simeq 0.086$), we calculate the optical force components for various complex values of the particle permittivity $\varepsilon_p$. In Table 1 we show the comparison of the forces obtained from the exact complex-angle Mie-scattering calculations and from the dipole-approximation equations (19) and (20), both for the *s*-polarized incident wave. Evidently, the agreement is very good, with deviations within a few percent caused by the accuracy of the dipole approximation.

Table 1. Comparison of radiation forces for a particle with the parameters (21) and different permittivities $\varepsilon_p$, *s*-polarized incident wave, calculated using: (i) the dipole approximation [23], Eqs. (19) and (20), and (ii) the exact complex-angle Mie theory (10)–(16).

| Force component | $\varepsilon_p = 2.25$ | | $\varepsilon_p = 15 + 0.14i$ | |
|---|---|---|---|---|
| | Dipole approximation | Complex-angle Mie theory | Dipole approximation | Complex-angle Mie theory |
| $F_x / P_0$ | $1.446 \cdot 10^{-4}$ | $1.420 \cdot 10^{-4}$ | $4.778 \cdot 10^{-3}$ | $4.863 \cdot 10^{-3}$ |
| $F_z / P_0$ | $-6.324 \cdot 10^{-2}$ | $-6.233 \cdot 10^{-2}$ | $-1.771 \cdot 10^{-1}$ | $-1.760 \cdot 10^{-1}$ |

| Force component | $\varepsilon_p = -5.65 + 0.75i$ | | $\varepsilon_p = -1$ | |
|---|---|---|---|---|
| | Dipole approximation | Complex-angle Mie theory | Dipole approximation | Complex-angle Mie theory |
| $F_x / P_0$ | $4.117 \cdot 10^{-1}$ | $4.164 \cdot 10^{-1}$ | $6.687 \cdot 10^{-3}$ | $6.334 \cdot 10^{-3}$ |
| $F_z / P_0$ | $-3.846 \cdot 10^{-1}$ | $-3.824 \cdot 10^{-1}$ | $4.300 \cdot 10^{-1}$ | $4.161 \cdot 10^{-1}$ |

Thus, we have shown that the complex-angle Mie theory results for the radiation forces are fully consistent with other approaches and approximations. Now we demonstrate an application of the proposed theory to the study of optical forces in evanescent fields. The most interesting situations occur in the case of *conducting* particles. For dielectric particles, the force $F_z$ is usually negative in the range of parameters considered [13,14,18,20,23] (see Fig. 4), i.e. attracts a particle towards the surface. It was suggested in [14] that the particle's conductivity can be a source of the positive $F_z$; here we consider how this effect can be evaluated using our complex-angle Mie theory.

We first consider a gold particle in water (Fig. 5a). This case is characterized by a significant imaginary part of the permittivity $\varepsilon_p = n_p^2$ [37,38], which likely promotes high absolute values of the optical forces. At the same time, the considerable difference in refraction indices $n_p$ and $n$ contributes to the oscillatory behavior of the curves. Note that $|F_x| > |F_z|$ for gold particles for almost the whole range of particle sizes. Furthermore, Figure 5a shows the possibility of *positive* $F_z$ for large enough particles $ka > 2$ and s-polarized illumination. This tendency becomes dominant for a "perfect metal" particle with $\varepsilon_p = -1$ (Fig. 5b), where the vertical force $F_z$ is always positive. The perfect-metal model qualitatively represents optical properties of some well-conducting metals at frequencies below the plasmon resonance [2]. All components of the normalized optical force exerted on conducting particles show a rather fast attenuation when the particle size increases, which can be attributed to the influence of absorption. The suppressed penetration of the radiation inside the particle and thus absence of the in-particle resonances is likely responsible for the fact that in Fig. 5b there are no oscillations, in contrast to the case of a dielectric particle (Fig. 4).

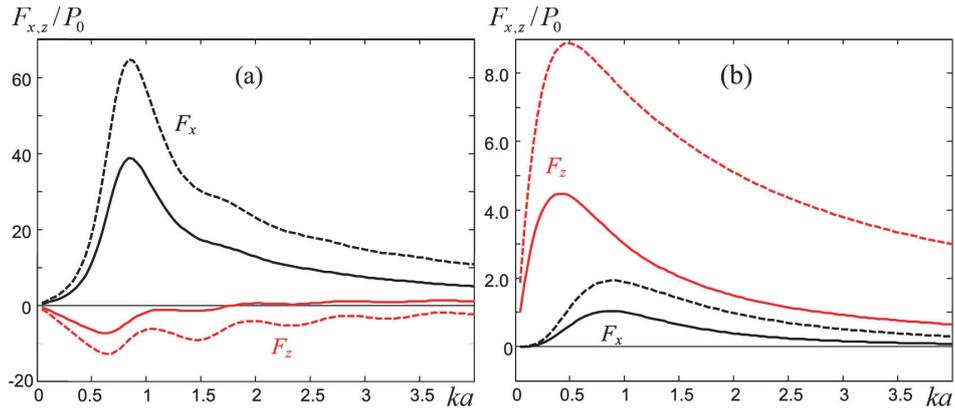

**Fig. 5.** Dimensionless radiation force components $F_x$ (black curves) and $F_z$ (red curves) versus the particle size parameter $ka$ for the *s*-polarized (solid curves) and *p*-polarized (dashed curves) incident wave. The parameters are the same as in Eq. (17) but with (a) $n = 1.33$, $n_p = 0.43 + 3.52i$ (gold particle in water at $\lambda = 650$ nm [3]) and (b) $n = 1$, $n_p = i$ ("perfect metal" particle).

Finally, we characterize the physical origin of the optical forces considered above. In all cases, $F_x > 0$, i.e., the force $F_x$ is directed along the field momentum $p_x$ (9). This enables us to associate the horizontal force with the surface energy flow of the evanescent field [the momentum component $p_y$ (9) vanishes in the case of *s*- or *p*-polarizations]. At $ka \ll 1$, the horizontal force grows as $F_x \propto a^6$ in Fig. 4 and in Fig. 5b, which is typical for the electromagnetic momentum action on particles with real polarizability in the dipole approximation [37,38]. In contrast, $F_x \propto a^3$ in Fig. 5a due to the non-zero imaginary part of the complex polarizability $\varepsilon a^3 \frac{\varepsilon_p - \varepsilon}{\varepsilon_p + 2\varepsilon}$ [22,23,37] [see Eq. (19)]. At the same time, the vertical force shows a characteristic gradient-force behavior: $F_z \propto a^3$ at $ka \ll 1$ [see Eq. (20)], which is not surprising since $p_z = 0$, and the force appears due to the inhomogeneous distribution of the energy density $w$ (9).

## 5. Conclusion

We have proposed a simple and efficient method for calculating the light scattering and the radiation force induced by an evanescent field interacting with a spherical particle. Our approach consists of a single complex-angle rotation applied to the standard Mie-scattering solutions. The results obtained by our method precisely coincide with those obtained within previous exact, but much more laborious and cumbersome, approaches [10–15] or, for the case of small particles, within the dipole-scattering approximation [19–23]. At the same time, the complex-angle Mie theory offers considerable advantages including a more transparent and time-saving procedure, use of well-elaborated calculation schemes and software codes.

We have illustrated the efficiency and applications of our approach by calculating the angular distributions of the far field scattered from the incident evanescent wave as compared to the known case of the propagating plane-wave incidence. Noteworthy, essentially the same code was used for both cases. Furthermore, we have examined the radiation forces exerted on various dielectric and metallic particles immersed in the evanescent field from a total-reflecting dielectric interface. All calculations were made for $s$ and $p$ linear polarizations of the incident wave and for different particle sizes. We have found that the vertical force can be repulsive for metallic particles with size comparable to the wavelength, in agreement with previous anticipations [14,15,19,20,23,24,31].

**Acknowledgements**

This work was supported by the European Commission (Marie Curie Action), ARO, JSPS-RFBR contract No. 12-02-92100, Grant-in-Aid for Scientific Research (S), MEXT Kakenhi on Quantum Cybernetics, and the JSPS via its FIRST program.

**Appendix A: Mie scattering formulas**

Here we collect formulas of the standard Mie theory. We mostly follow [4] but modify and adapt the Mie equations with special attention to the scattering near field and to the radial components which are typically omitted. As usual, it is assumed that a spherical particle of radius $a$ (placed at the origin) with electromagnetic parameters $\varepsilon_p$, $\mu_p$, $n_p = \sqrt{\varepsilon_p \mu_p}$ scatters the incident $z$-propagating plane wave (1) in a medium with parameters $\varepsilon$, $\mu$, and $n = \sqrt{\varepsilon\mu}$. Using spherical coordinates $(\theta, \phi, r)$ introduced with respect to the $(x, y, z)$ coordinates (see Fig. 1), the scattering is described with respect to planes determined by the azimuthal cross-sections $\phi = \text{const}$. In this manner, the in-plane and out-of-plane components of the amplitudes of the incident field (1) are re-defined as

$$\begin{pmatrix} \tilde{E}_\parallel \\ \tilde{E}_\perp \end{pmatrix} = \begin{pmatrix} \cos\phi & \sin\phi \\ -\sin\phi & \cos\phi \end{pmatrix} \begin{pmatrix} E_\parallel \\ E_\perp \end{pmatrix}, \quad \begin{pmatrix} \tilde{H}_\parallel \\ \tilde{H}_\perp \end{pmatrix} = \sqrt{\frac{\varepsilon}{\mu}} \begin{pmatrix} \cos\phi & \sin\phi \\ -\sin\phi & \cos\phi \end{pmatrix} \begin{pmatrix} -E_\perp \\ E_\parallel \end{pmatrix} \tag{A1}$$

In spherical coordinates, the components of the scattered fields, $\mathbf{E}^s$ and $\mathbf{H}^s$, are calculated via the following expansions:

$$E_\theta^s = \tilde{E}_\parallel \frac{1}{r} \sum_{\ell=1}^{\infty} A_\ell \left( i a_\ell \xi_\ell' \tau_\ell - b_\ell \xi_\ell \pi_\ell \right), \quad H_\theta^s = -\tilde{E}_\perp \frac{1}{r} \sqrt{\frac{\varepsilon}{\mu}} \sum_{\ell=1}^{\infty} A_\ell \left( i b_\ell \xi_\ell' \tau_\ell - a_\ell \xi_\ell \pi_\ell \right),$$

$$E_\phi^s = \tilde{E}_\perp \frac{1}{r} \sum_{\ell=1}^{\infty} A_\ell \left( b_\ell \xi_\ell \tau_\ell - i a_\ell \xi_\ell' \pi_\ell \right), \quad H_\phi^s = \tilde{E}_\parallel \frac{1}{r} \sqrt{\frac{\varepsilon}{\mu}} \sum_{\ell=1}^{\infty} A_\ell \left( i b_\ell \xi_\ell' \pi_\ell - a_\ell \xi_\ell \tau_\ell \right),$$

$$E_r^s = \tilde{E}_\parallel \sin\theta \frac{1}{r^2} \sum_{\ell=1}^{\infty} A_\ell \ell(\ell+1) i a_\ell \xi_\ell \pi_\ell \;,\quad H_r^s = -\tilde{E}_\perp \sin\theta \frac{1}{r^2} \sqrt{\frac{\varepsilon}{\mu}} \sum_{\ell=1}^{\infty} A_\ell \ell(\ell+1) i b_\ell \xi_\ell \pi_\ell \;. \quad (A2)$$

Here $A_\ell = i^\ell \dfrac{2\ell+1}{\ell(\ell+1)}$, and each term in the sums describes a certain order of multipole radiation, namely, the $a_\ell$- and $b_\ell$-terms represent the electric and magnetic $2^\ell$-poles located at the origin. The radial and polar dependences of the solutions (A2) are contained in the functions

$$\xi_\ell(kr) = kr\, h_\ell^{(1)}(kr)\,,\quad \xi_\ell'(kr) = \frac{d\left[kr\, h_\ell^{(1)}(kr)\right]}{d(kr)}\,,$$

$$\pi_\ell(\cos\theta) = \frac{P_\ell^1(\cos\theta)}{\sin\theta}\,,\quad \tau_\ell(\cos\theta) = \frac{dP_\ell^1(\cos\theta)}{d\theta}\,, \quad (A3)$$

where $h_\ell^{(1)}(u)$ are the spherical Hankel functions and $P_\ell^1(u)$ are the adjoint Legendre polynomials (well-defined in the complex domain). The scattering coefficients $a_\ell(\chi)$ and $b_\ell(\chi)$ depend on the dimensionless particle radius $\chi = ka$ and are given by

$$a_\ell = \frac{m_\varepsilon \psi_\ell(m\chi)\psi_\ell'(\chi) - m_\mu \psi_\ell(\chi)\psi_\ell'(m\chi)}{m_\varepsilon \psi_\ell(m\chi)\xi_\ell'(\chi) - m_\mu \xi_\ell(\chi)\psi_\ell'(m\chi)}\,,\quad b_\ell = \frac{m_\mu \psi_\ell(m\chi)\psi_\ell'(\chi) - m_\varepsilon \psi_\ell(\chi)\psi_\ell'(m\chi)}{m_\mu \psi_\ell(m\chi)\xi_\ell'(\chi) - m_\varepsilon \xi_\ell(\chi)\psi_\ell'(m\chi)}\,. \quad (A4)$$

Here the following parameters and functions are used:

$$m_\varepsilon = \sqrt{\frac{\varepsilon_p}{\varepsilon}}\,,\quad m_\mu = \sqrt{\frac{\mu_p}{\mu}}\,,\quad m = \frac{n_p}{n}\,,$$

$$\psi_\ell(u) = u\, j_\ell(u)\,,\quad \psi_\ell'(u) = \frac{d\left[u\, j_\ell(u)\right]}{du}\,, \quad (A5)$$

where $j_\ell(u)$ are the spherical Bessel functions.

Finally, the Cartesian components of the scattered field are obtained via the standard rotational transformation connecting Cartesian and spherical coordinates:

$$\mathbf{E}^s = \begin{pmatrix} E_x^s \\ E_y^s \\ E_z^s \end{pmatrix} = \hat{R}(\theta,\phi) \begin{pmatrix} E_\theta^s \\ E_\phi^s \\ E_r^s \end{pmatrix}\,,\quad \mathbf{H}^s = \begin{pmatrix} H_x^s \\ H_y^s \\ H_z^s \end{pmatrix} = \hat{R}(\theta,\phi) \begin{pmatrix} H_\theta^s \\ H_\phi^s \\ H_r^s \end{pmatrix},$$

$$\hat{R}(\theta,\phi) = \hat{R}_z(-\phi)\hat{R}_y(-\theta) = \begin{pmatrix} \cos\theta\cos\phi & -\sin\phi & \sin\theta\cos\phi \\ \cos\theta\sin\phi & \cos\phi & \sin\theta\sin\phi \\ -\sin\theta & 0 & \cos\theta \end{pmatrix}. \quad (A6)$$

Equations (1), (A1)–(A6) represent the standard Mie scattering solution establishing the linear relation (12) between the scattered and incident ($z$-propagating plane wave) fields.